\documentclass[fleqn,10pt]{wlscirep}
\usepackage[utf8]{inputenc}
\usepackage[T1]{fontenc}
\title{Metasurface Sensing Difference in Waveforms at the Same Frequency with Reduced Power Level}

\author[1]{Mizuki Tanikawa}
\author[1]{Daiju Ushikoshi}
\author[1]{Kosei Asano}
\author[2]{Kenichiro Sanji}
\author[2]{Masakazu Ikeda}
\author[1]{Daisuke Anzai}
\author[1,3,*]{Hiroki Wakatsuchi}
\affil[1]{Department of Electrical and Mechanical Engineering, Graduate School of Engineering, Nagoya Institute of Technology, Nagoya, Aichi, 466-8555, Japan}
\affil[2]{Research Department 23, Research \& Development Department 2, SOKEN, INC., Nisshin, Aichi, 470-0111, Japan}
\affil[3]{Precursory Research for Embryonic Science and Technology (PRESTO), Japan Science and Technology Agency (JST), Kawaguchi, Saitama, 332-0012, Japan}

\affil[*]{wakatsuchi.hiroki@nitech.ac.jp}

\begin{abstract}
We numerically demonstrate a new type of waveform-selective metasurface that senses the difference in incoming waveforms or pulse widths at the same frequency. Importantly, the proposed structure contains precise rectifier circuits that, compared to ordinary schottky diodes used within old types of structures, rectify induced electric charges at a markedly reduced input power level depending on several design parameters but mostly on the gain of operational amplifiers. As a result, a waveform-selective absorbing mechanism related to this turn-on voltage appears even with a limited signal strength that is comparable to realistic wireless signal levels. In addition, the proposed structure exhibits a noticeably wide dynamic range from -30 to 6 dBm, compared to a conventional structure that operated only around 0 dBm. Thus, our study opens up the door to apply the concept of waveform selectivity to a more practical field of wireless communications to control different small signals at the same frequency. 
\end{abstract}
\begin{document}

\flushbottom
\maketitle
%
%
\thispagestyle{empty}

In modern society our daily life benefits from various wireless communication devices including smartphones, Bluetooth/WiFi/IoT (Internet of Things) devices, broadcasting antennas for television/radio, radar systems for weather forecasting and GPS (global positioning system) satellites. On one hand, there is strong demand on developing more advanced wireless communication technologies or services to further enhance the quality of our life. On the other hand, this continuous development escalates a potential risk for communication devices to be exposed to other wireless signals unnecessarily, leading to unexpected electromagnetic interference issues \cite{CCemcBook}. These issues are especially serious in the ISM (industrial, scientific and medical) bands, which are standardized to be readily used without strict licence procedures. For instance, 13.5 MHz and 2.4 GHz are well known to be used for RFID and WiFi technologies, respectively. 

A conventional solution to these interference issues is deploying absorbent materials that effectively dissipate the energy of an incoming wave at a designed frequency (or band) and thus suppress the magnitude of the scattering wave to an acceptable level \cite{knott2006radar}. Although a traditional type of absorber was relatively bulky to the wavelength of an incident wave \cite{salisbury1952absorbent}, absorbers composed of artificially engineered periodic structures, or the so-called metamaterials/metasurfaces \cite{pendryENG1, pendryMNG, smithDNG1D, smithDNG2D2, EBGdevelopment, THzActiveMTMpadilla, sievenpiper2010, yu2011light}, can be designed in a limited space with subwavelength thicknesses and light weights, thereby enhancing their applicability to various situations and successfully mitigating electromagnetic interference issues \cite{MunkBook, watts2012metamaterial, li2017nonlinear, mtmAbsPRLpadilla, mtmAbsOEpadilla, ultraThinAbs, My1stAbsPaper, aplNonlinearMetasurface, li2017high}. However, electromagnetic interference becomes more complicated once the spectrum of a wireless communication signal overlaps that of other signal/noise. This is because if the incoming spectrum is fixed, ordinary materials similarly respond to different waves and thus are incapable of distinguishing one signal from another. As opposed to these conventional materials, however, circuit-based metasurface absorbers were recently demonstrated to be capable of absorbing a particular type of incoming wave even at the same frequency by sensing its waveform or pulse width \cite{wakatsuchi2013waveform, eleftheriades2014electronics, wakatsuchi2015waveformSciRep, vellucci2018towards, wakatsuchi2019waveform, vellucci2019waveform}. Such a waveform-selective mechanism was thus expected to give us an additional degree of freedom to control electromagnetic waves/signals and solve electromagnetic interference issues occurring at the same frequency. A remaining issue here, however, is that due to the turn-on voltage of the diodes integrated with the structures, any of waveform-selective metasurfaces reported to date required an extremely large input power level, compared to the strength of most wireless communication signals \cite{goldsmith2005wireless}. Hence, this issue hindered fully exploiting the concept of waveform selectivity in realistic wireless communication environment. 

For this reason, we propose a new type of waveform-selective metasurface that is numerically demonstrated to operate at a low power level but still selectively absorbs a particular waveform at the same frequency. This is achieved by replacing conventional diodes with precise rectifier circuits (PRCs) \cite{stout1976handbook} that markedly lower turn-on voltage depending on the gain of operational amplifiers (op-amps) used.

\section*{Structure and waveform-selective mechanism}
The waveform-selective metasurface demonstrated in this numerical study consisted of metallic patches (1,600 $\times$ 1,600 mm$^2$), a dielectric substrate (Rogers3003, 300 mm thickness) and a ground plane (Fig.\ \ref{model}a). The periodicity of the structure was set to a large value (1,700 mm), although entire dimensions can be readily scaled down by introducing lumped capacitors between metallic patches to lower the resonant frequency even with a small periodicity. Additionally, each gap between conducting patches was connected by four diodes that worked as a diode bridge (Figs.\ \ref{model}b and c). Under these circumstances, the diode bridge converts the incoming spectrum to an infinite set of frequency components, although most of the energy appears at zero frequency as the waveform is fully rectified \cite{wakatsuchi2013waveform, wakatsuchi2015waveformSciRep, wakatsuchi2019waveform}. Therefore, by connecting a capacitor to a resistor in parallel within the diode bridge (Figs.\ \ref{model}b and c) \cite{wakatsuchi2013waveform}, the energy of the incoming pulsed wave can be firstly stored within the structure and dissipated at the resistor before a next pulse comes in. However, this capacitor-based waveform-selective metasurface lowers its absorbing performance when the incoming waveform changes to long pulse or continuous wave (CW), because the capacitor used is fully charged up. Other types of waveform-selective metasurfaces are seen in our previous studies \cite{wakatsuchi2015waveformSciRep, wakatsuchi2015time}. 

\section*{Simulation method}
\begin{figure}[!t]
\centerline{\includegraphics[width=0.7\columnwidth]{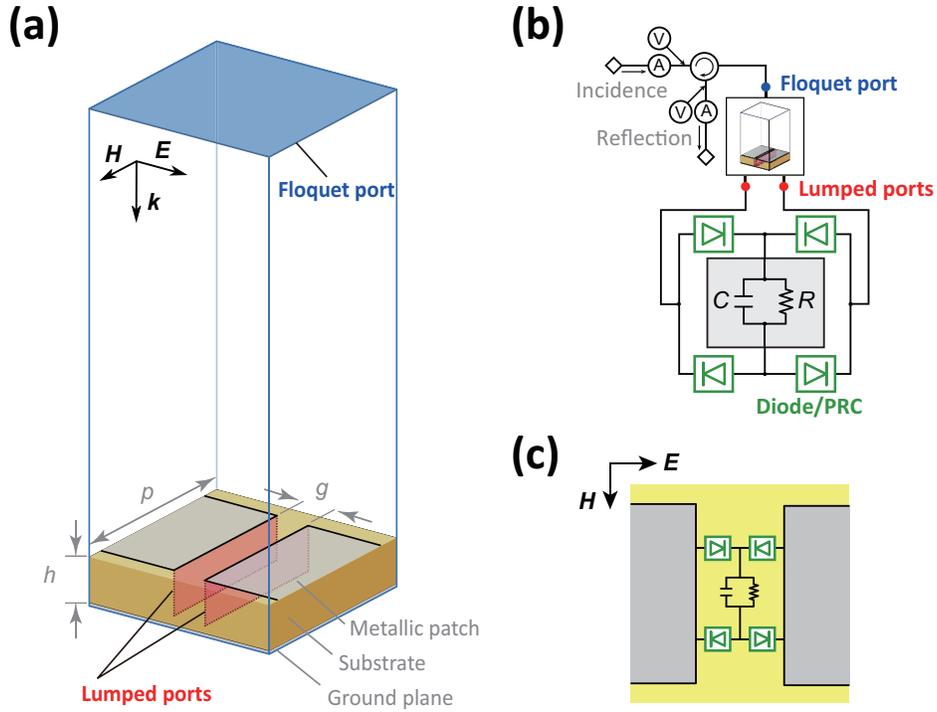}}
\caption{(a) Electromagnetic model, (b) circuit schematic diagram and (c) equivalent electromagnetic structure. Additional circuit components deployed inside a diode bridge. Design parameters are given in Table \ref{tab:dimensions}, and PRC is seen in Fig.\ \ref{PRC}a. }
\label{model}
\end{figure}

This structure was numerically tested by using a co-simulation method integrating an electromagnetic simulator with a circuit simulator (ANSYS Electronics Desktop R18.1.0) \cite{wakatsuchi2013waveform}. Compared to ordinary electromagnetic simulation methods \cite{jin2015finite, taflove2005computational, TLMbook1}, the co-simulation method significantly reduced a total simulation time at the expense of fully visualizing spatial electromagnetic field distribution \cite{wakatsuchi2015fieldVisualisation}, since eventually all of simulation results were obtained in circuit simulations (Fig.\ \ref{model}b). In the co-simulation method, a single periodic unit cell was modelled using periodic boundaries to effectively represent an infinite array of periodic structure (see Fig.\ \ref{model}a). An incident wave was generated from a Floquet port deployed on the top of the entire analysis space. The frequency was changed from 20 to 50 MHz with the power level varied from -50 to 0 dBm. Unlike our past studies \cite{wakatsuchi2015waveformJAP} where only ``one'' lumped port was used to model circuit components (i.e., a diode bridge, resistor and capacitor) between two conducting patches and to effectively assume one of the conductor edges as a ground, the model of Fig.\ \ref{model}a had ``two'' lumped ports between a ground plane and patches (see also Fig.\ \ref{model}b). This is because the new rectifying circuits proposed later in this study contained additional grounds that could not be electrically connected to the above-mentioned ground.  

\begin{table}[h!]
\caption{\label{tab:dimensions} Design parameters used for a periodic unit cell of the waveform-selective metasurface drawn in Fig.\ \ref{model}. }
\begin{center}
\begin{tabular}{cc||cc}
\hline 
Variables & Values & Variables & Values\\
\hline 
\hline 
$p$ & 1,700 mm & $C$ & 100 nF\\
$h$ & 300 mm & $R$ & 100 k$\Omega$\\
$g$ & 100 mm & \\
\hline 
\end{tabular}
\end{center}
\end{table}

\section*{Precise rectifier circuit}
\begin{figure}[!t]
\centerline{\includegraphics[width=0.6\columnwidth]{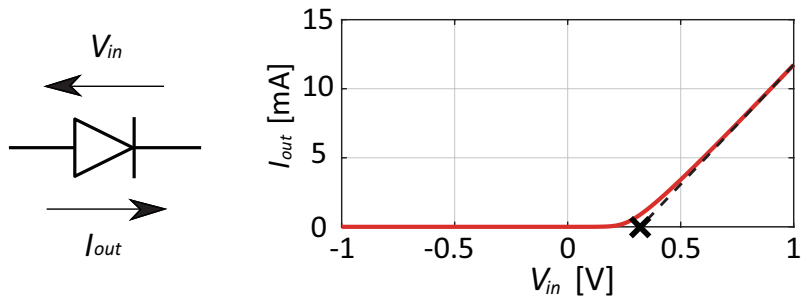}}
\caption{$I$-$V$ characteristics of schottky diode used in this study (Broadcom, HSMS2860). The black cross represents its turn-on voltage.}
\label{circuit}
\end{figure}

Practically, commercial schottky diodes have turn-on voltage of a few hundred mV or so. For example, Fig.\ \ref{circuit} shows the relationship between voltage and current, namely, $I$-$V$ curve of a schottky diode (Broadcom, HSMS2860 but without parasitic junction capacitance $C_{J0}$ and series resistance $R_S$ for the sake of simplicity). By approximating this curve by a straight line at a sufficiently large voltage (e.g., 1 V) and reading its intersection with the horizontal axis (as marked by the black cross in Fig.\ \ref{circuit}), turn-on voltage $V_{on}$ of the diode is estimated to be around 0.3 V. 

\begin{figure}[!t]
\centerline{\includegraphics[width=\columnwidth]{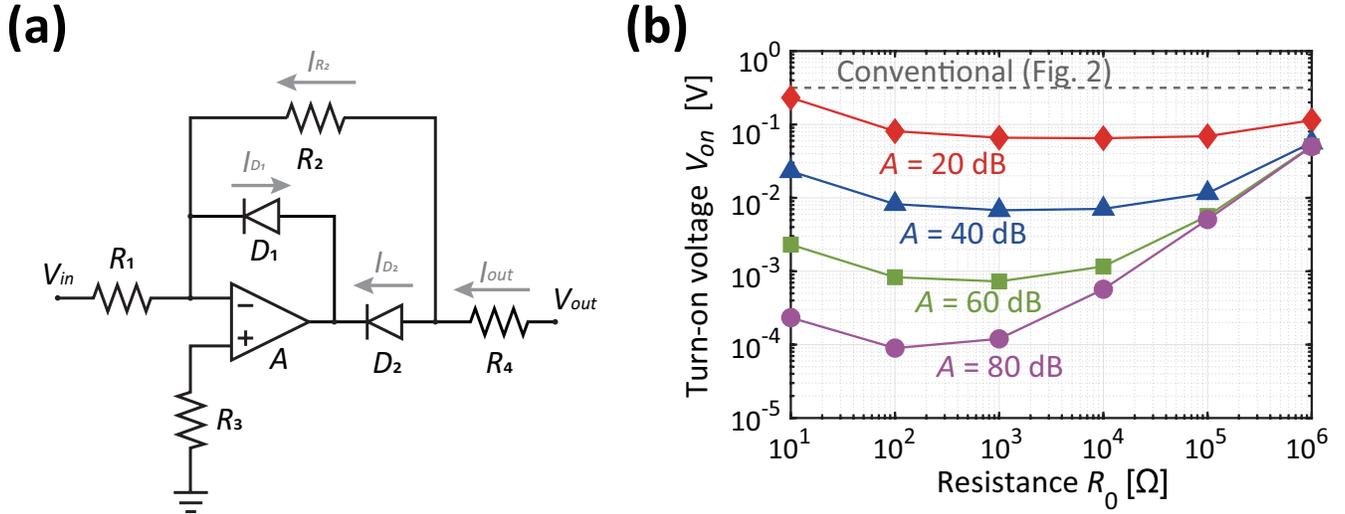}}
\caption{(a) Proposed PRC and (b) its turn-on voltage with resistances set to $R_1=R_2=R_3/2=R_0$ and $R_4=50$ $\Omega$. For the top two PRCs of Figs.\ \ref{model}b and c, the orientations of $D_1$ and $D_2$ were changed, while those for the bottom two remained the same as the ones shown above. }
\label{PRC}
\end{figure}

In this study, $V_{on}$ was decreased by alternatively using the PRC drawn in Fig.\ \ref{PRC}a \cite{stout1976handbook}. Basically, this circuit took the form of an inverse amplifier circuit and contained one ideal op-amp, four resistors and two diodes (the same model as Fig.\ \ref{circuit}). In general, output voltage $V_{out}$ of this entire circuit is associated with input voltage $V_{in}$ as 
\begin{equation}
	V_{out} =  \begin{cases}
		-R_{2}(V_{in}-V_{on})/R_{1} & (V_{in}>V_{on}) \\
		0 &(V_{in}<V_{on})
	,\end{cases} 
\end{equation}
where $R_1$ and $R_2$ are resistances used in the circuit (see Fig.\ \ref{PRC}a).  Ideally, $V_{on}$ is 
\begin{equation}
	V_{on} = 	V_{0}/A,
\end{equation}
where $V_{0}$ is the turn-on voltage of the diodes included in the PRC, and $A$ represents the gain of the op-amp. Hence, $V_{on}$ is inversely proportional to $A$. 

\begin{figure}[!t]
\centerline{\includegraphics[width=0.8\columnwidth]{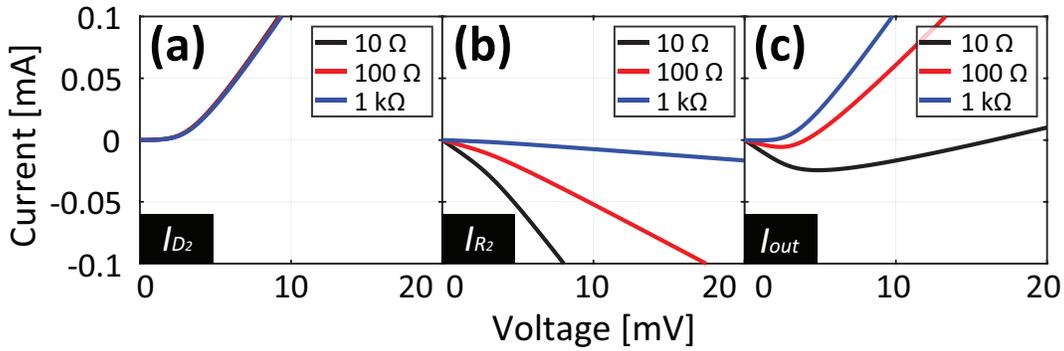}}
\caption{Currents at (a) $D_2$, (b) $R_2$ and (c) $R_4$ shown in Fig.\ \ref{PRC} with small $R_0$. }
\label{smallR}
\end{figure}

However, this turn-on voltage is also influenced by resistances as demonstrated in Fig.\ \ref{PRC}b, where $R_1$, $R_2$ and $R_3/2$ are determined by $R_0$, while $R_4$ is fixed at 50 $\Omega$. For instance, with small $R_{0}$ the current flowing into diode $D_2$ of Fig.\ \ref{PRC} is almost independent of $R_0$ (Fig.\ \ref{smallR}a). However, the current at $R_2$ varies as the resistor is effectively shorted (Fig.\ \ref{smallR}b). Since these two currents determine the output current based on Kirchhoff's current law, the turn-on voltage increases by using a small resistance value such as 10 $\Omega$ compared to 1 k$\Omega$ (Fig.\ \ref{smallR}c). Note that these current curves are changed by the gain of the op-amp so that the turn-on voltage is still inversely proportional to $A$ (Fig.\ \ref{PRC}b).  

\begin{figure}[!t]
\centerline{\includegraphics[width=\columnwidth]{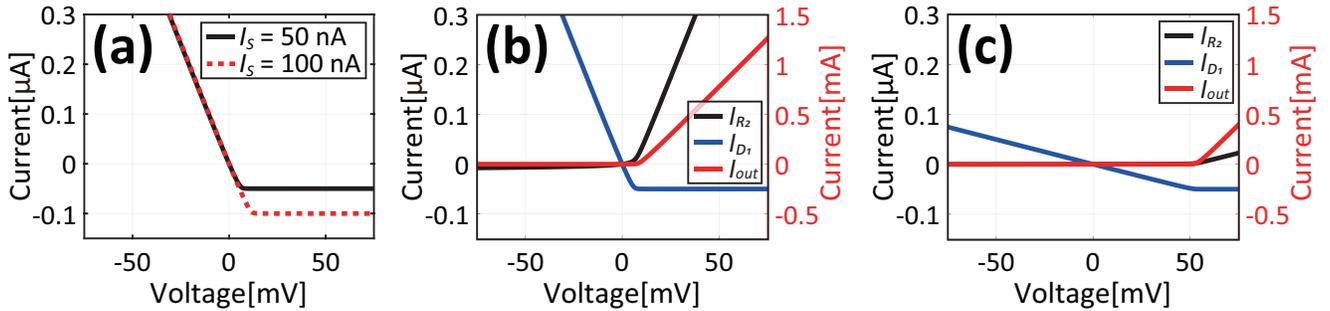}}
\caption{(a) Current at $D_1$ shown in Fig.\ \ref{PRC} with various saturation current values. Currents at $D_1$, $R_2$ and $R_4$ with (b) $R_0$ = 100 k$\Omega$ and (c) $R_0$ = 1 M$\Omega$. }
\label{largeR}
\end{figure}

With extremely large $R_0$ such as $R_0$ = 1 M$\Omega$, the turn-on voltage increases again. Although an ideal diode does not permit electric charges to enter from its anode to cathode, a more realistic diode has saturation current, which varies the amount of the current reversely coming into $D_1$ of Fig.\ \ref{PRC}a as drawn in Fig.\ \ref{largeR}a, where saturation current $I_S$ is set to two different values. When $I_S$ saturates, $R_2$ starts increasing its current, which relates to the output current as well (Fig.\ \ref{largeR}b). Importantly, the larger $R_0$, the larger $V_{on}$ becomes (cf.\ Fig.\ \ref{largeR}c). Also, these characteristics are independent of the gain of the op-amp so that with large $R_0$, $V_{on}$ becomes not only larger but also independent of $A$ (see around $R_0$ = 1 M$\Omega$ in Fig.\ \ref{PRC}b). 

Based on these results, circuit parameters need to be properly designed, otherwise the operating power level of a waveform-selective metasurface would not be improved. In the following part of this study, $R_{1}$, $R_{2}$, $R_{3}$ and $R_{4}$ are, respectively, set to 10 k$\Omega$, 10 k$\Omega$, 5 k$\Omega$ and 50 $\Omega$ as default values, while $A$ is basically fixed at 40 dB (Table \ref{tab:PRC}).  

\begin{table}[h!]
\caption{\label{tab:PRC} Default values used for precise rectifier circuits. }
\begin{center}
\begin{tabular}{cc||cc}
\hline 
Variables & Values & Variables & Values \\
\hline 
\hline 
$A$ & 40 dB & $R_3$ & 5 k$\Omega$\\
$R_1$, $R_2$ & 10 k$\Omega$ & $R_4$ & 50 $\Omega$\\
\hline 
\end{tabular}
\end{center}
\end{table}

In addition, the orientations of $D_1$ and $D_2$ are changed for the top two PRCs of Fig.\ \ref{model}b, while those for the bottom two remain the same as the ones shown in Fig.\ \ref{PRC}a. This is because electric charges coming into an op-amp change their sign. Therefore, unless the diodes of the next PRC are aligned for another direction, the electric charges are not allowed to reach an adjacent conducting patch (also, their sign is not restored).

\section*{Simulated absorbing performance}

\begin{figure}[!t]
\centerline{\includegraphics[width=0.8\columnwidth]{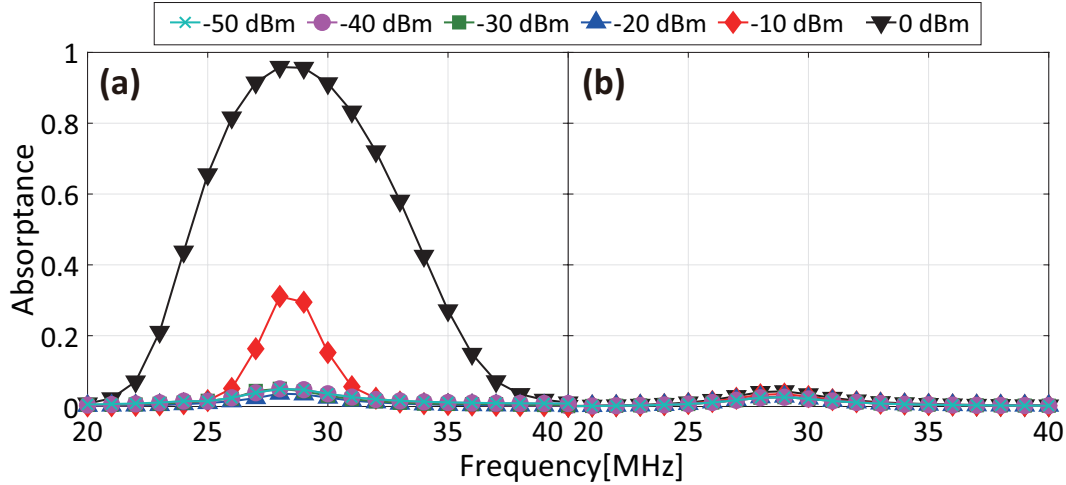}}
\caption{Absorptance of a conventional capacitor-based waveform-selective metasurface using schottky diodes for (a) 5-$\mu$s short pulse and (b) CW.}
\label{prevC}
\vspace{5mm}
\end{figure}

To evaluate improvement on the operating power of the proposed structure, we firstly simulated a capacitor-based waveform-selective metasurface using conventional commercial schottky diodes as shown in Fig.\ \ref{prevC}. This figure demonstrates that the structure was incapable of strongly absorbing any waveform within the power range between -50 and -20 dBm. Then the capacitor-based waveform-selective metasurface enhanced its absorptance for a short pulse by increasing the power level to 0 dBm, which, however, exceeds the strength of ordinary wireless communication signals \cite{goldsmith2005wireless}. We also note that this absorbing mechanism was consistent with the theory mentioned above as well as with previous studies despite the difference in operating frequencies\cite{wakatsuchi2015waveformJAP}. Another point here is that the difference in absorptances of a short pulse and a CW is larger than that in our previous reports. This is because our structure used relatively a large time constant, which increased the difference between its response to a short pulse and that to a CW and thus contributed to enhancing the waveform-selective performance. 

\begin{figure}[!t]
\centerline{\includegraphics[width=0.8\columnwidth]{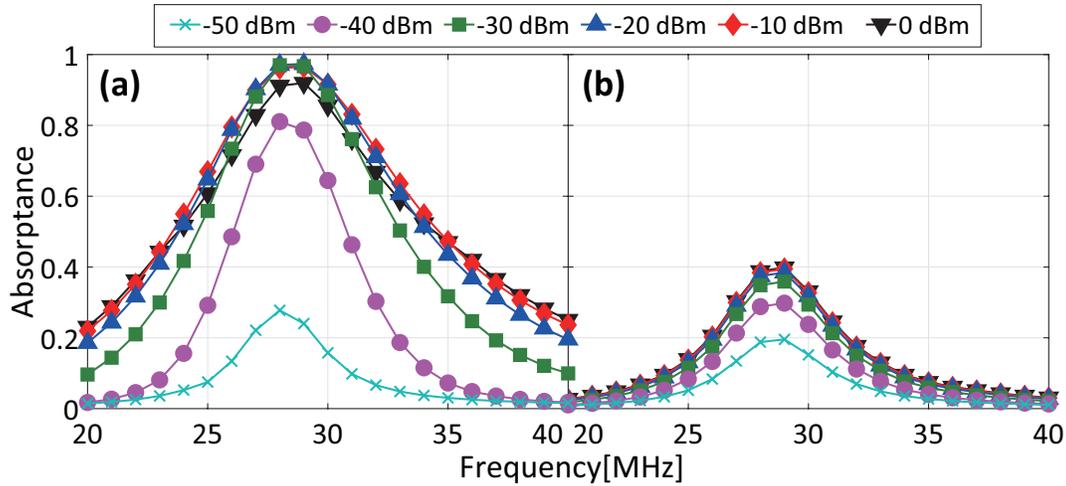}}
\caption{Absorptance of a proposed capacitor-based waveform-selective metasurface using PRCs for (a) 5-$\mu$s short pulse and (b) CW.}
\label{newC}
\vspace{5mm}
\end{figure}

Secondly, this structure was simulated using the proposed PRCs instead of the conventional diodes. As plotted in Fig.\ \ref{newC}, a waveform-selective absorbing mechanism appeared even if the input power was set to merely -40 dBm. The difference between the absorptance for a short pulse and that for a CW was maximized within the power range between -30 and -10 dBm. For instance, the capacitor-based waveform-selective metasurface absorbed 97.3 $\%$ of a short pulse and 38.5 $\%$ of a CW at the same frequency of 29 MHz. We also noticed that the gap between the pulse absorptance and the CW absorptance was reduced to about 60 $\%$ (with -20 dBm), compared to approximately 90 $\%$ difference seen in Fig.\ \ref{prevC} (with 0 dBm). This is probably because electric charges rectified by the PRCs were dissipated by not only resistors but also the PRCs that contained additional resistors. 

\begin{figure}[t]
\centerline{\includegraphics[width=0.4\columnwidth]{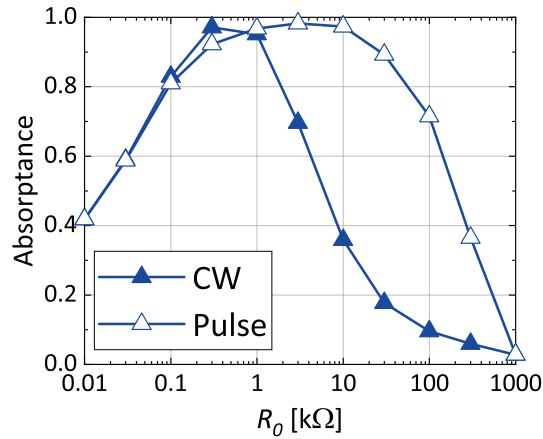}}
\caption{$R_0$ dependence of absorptance of capacitor-based waveform-selective metasurface using PRCs. The frequency and input power were fixed at 29 MHz and -30 dBm, respectively. }
\label{rDep}
\end{figure}

Regarding this point, Fig.\ \ref{rDep} shows how $R_0$ related to the absorptances of the capacitor-based waveform-selective metasurface using the PRCs. As seen in this figure, where the input frequency and power were fixed at 29 MHz and -30 dBm, respectively, the CW absorptance was found to be suppressed by increasing $R_0$ from 10 k$\Omega$. At the same time, however, this led to lowering the short-pulse absorptance as well. This is because entirely the turn-on voltage of the PRCs was increased as explained in Figs.\ \ref{PRC} and \ref{largeR}. Therefore, the difference in the absorptances needs to be increased by properly designing $R_0$. In addition, this absorptance gap may be enlarged by independently adjusting the resistive components of the PRCs, although our study set $R_1$, $R_2$ and $R_3$ to be $R_1=R_2=R_3/2$ for the sake of simplicity. Moreover, Fig.\ \ref{rDep} also shows that the CW absorptance approached the short-pulse absorptance by decreasing $R_0$ from 10 k$\Omega$. This trend is also explained by the turn-on voltage characteristics of the PCRs (refer to Figs.\ \ref{PRC} and \ref{smallR} again). We noticed that the short-pulse absorptance became slightly smaller than the CW absorptance for $R_0<$ 1k$\Omega$. This minor gap is possibly due to the small difference between the frequency spectra of both waveforms (i.e., the short pulse contained the oscillating frequency component mostly but also other minor components). 

\begin{figure}[t]
\centerline{\includegraphics[width=0.8\columnwidth]{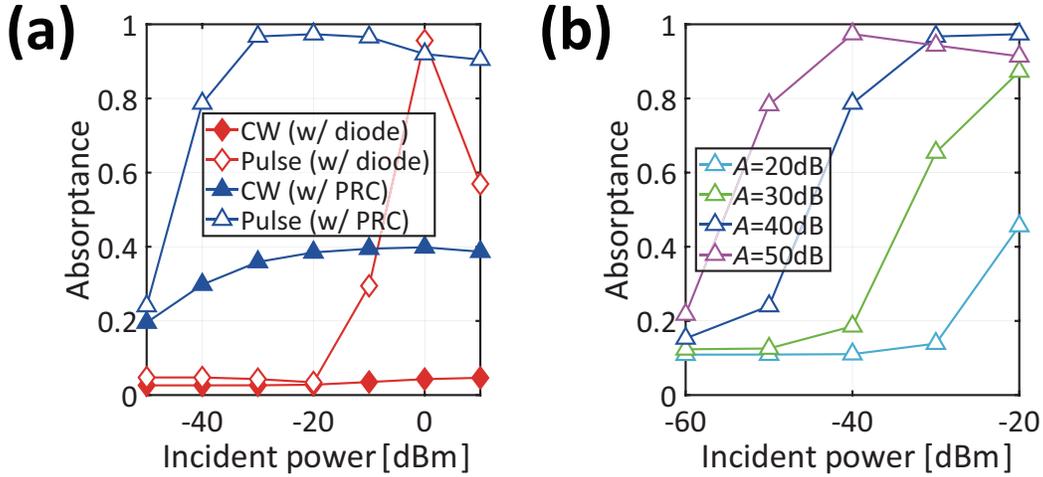}}
\caption{Power dependence of absorptance of (a) capacitor-based waveform-selective metasurfaces using either schottky diodes or PRCs with $A$ = 40 dB and frequency fixed at 29 MHz. (b) Power dependence of pulse absorptance of the proposed capacitor-based waveform-selective metasurface using PRCs with various gains.}
\label{pDep}
\end{figure}

The improvement on the operating power level of the proposed structure is more clearly seen in Fig.\ \ref{pDep}a, which plots the absorptances of the two structures as a function of input power (with frequency fixed at 29 MHz). According to this result, the conventional structure required 0 dBm input power to achieve almost 100 $\%$ absorptance. However, our proposed structure reduced this power level to -30 dBm or so. Note that this improvement (about 30 dB) almost matched the gain of the op-amp, which was set to 40 dB in this study, although a minor discrepancy appeared as the turn-on voltage of PRC was slightly worse than expected from the gain (refer to Fig.\ \ref{PRC}). However, our results still support that waveform-selective metasurfaces can operate at a markedly reduced power level by lowering turn-on voltage of rectifier circuits, which was achieved through use of PRCs. Additionally, Fig.\ \ref{pDep}b shows the short-pulse absorptance of the capacitor-based waveform-selective metasurface using PRCs with various gain values. This figure clearly demonstrates that the turn-on voltage is improved proportionally to gain $A$. 

\begin{figure}[t]
\centerline{\includegraphics[width=0.9\columnwidth]{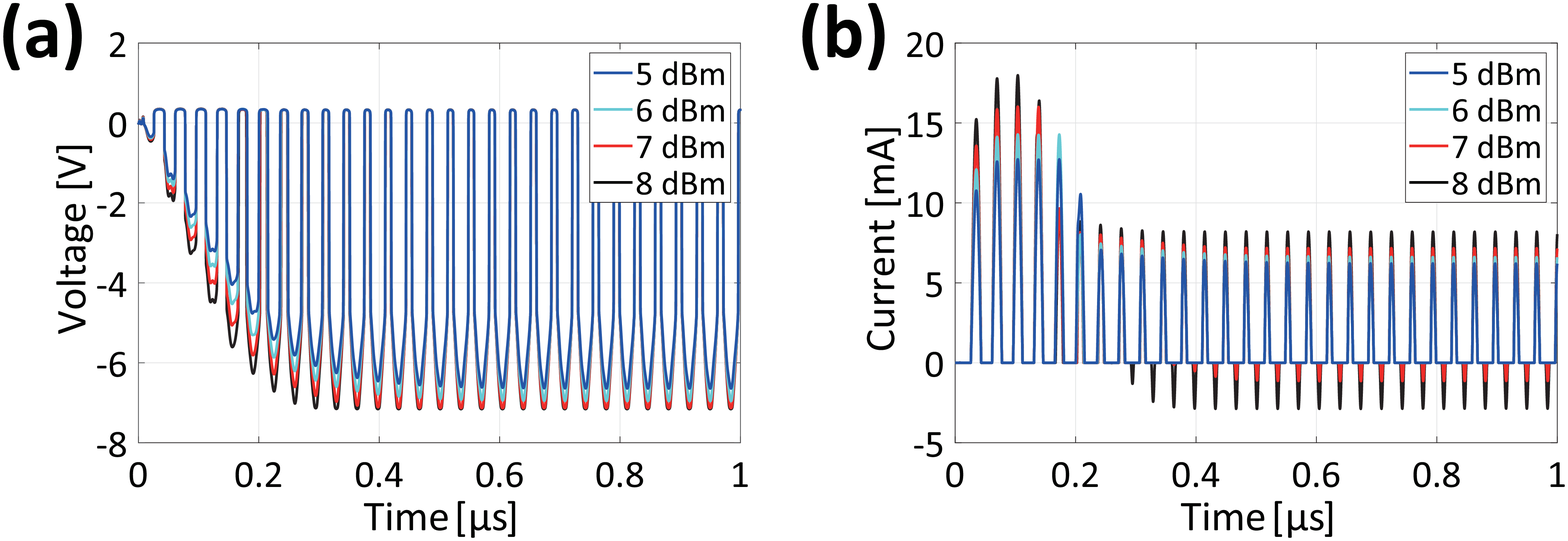}}
\caption{(a) Voltage and (b) current of $D_2$ used as the top two PRCs in Figs.\ \ref{model}b and c.}
\label{Vbv}
\end{figure}

Moreover, Fig.\ \ref{pDep}a indicates that our new structure not only lowered its operating power level but also improved a dynamic range of waveform-selective absorbing mechanism. Specifically, our structure showed almost 100 $\%$ absorptance between -30 dBm and -10 dBm, which was limited to only around 0 dBm in the conventional structure (compare the open triangles of Fig.\ \ref{pDep}a to the open rhombuses). In the conventional structure, basically the dynamic range is limited by the turn-on voltage of diodes as well as their break-down voltage. In our proposed structure, the diodes included in PRCs were found to operate for a wide range of input power level, as op-amps adjusted the amount of current to come into anodes of diodes. In the case of Fig.\ \ref{PRC}a, for example, the op-amp amplified the input voltage only to turn on $D_2$ so that the voltage across $D1$ did not reach its break-down voltage. In some cases, however, diodes were found to reach their break-down voltage as seen in Fig.\ \ref{Vbv}, which plots the voltage and current of diode $D_2$ whose cathodes were oriented to the output (i.e., the same configuration as the one used for the top two PRCs of Fig.\ \ref{model}b). This figure shows that with an input power larger than 6 dBm the voltage across $D_2$ became -7 V, which was its break-down voltage (Fig.\ \ref{Vbv}a). As a result, its current flowed both forwardly and reversely (Fig.\ \ref{Vbv}b). Nonetheless, the use of these new rectifying circuits led to a broader dynamic range from -30 dBm to 6 dBm than the one seen in the conventional structure. 

\section*{Discussion}

The improvement on the operating power level relates not only to the gain of op-amps and the resistive component of PRCs but also to other factors including frequency, as practically op-amps have frequency-dependent gain and thus more strongly/weakly amplify incoming electric charges depending on frequency. Although our study used PRCs, conventional schottky diodes may be replaced by other rectifier circuits, as long as induced electric charges are properly controlled to enter the internal circuit components. In particular, biased transistors can be alternatively used as amplifier circuit elements instead of op-amps. Many MOSFETs can work at a GHz range including 2.4 GHz (i.e., one of the ISM bands). For a higher frequency such as the millimetre wave band, the op-amps used in our study need to be replaced by more advanced amplifier circuit components (e.g., Microsemi, MMA025AA). In this case, potentially waveform selectivity can be achieved in the millimetre-wave band, although possibly more improvement/modification is required for the entire rectifier circuit design. Our waveform-selective metasurface was designed to operate around 30 MHz, where many existing commercial op-amps are expected to work (e.g., Texas Instruments provides the op-amps (OPA858) that have a gain larger than 40 dB in this frequency range). Although this study only provided numerical simulations, we emphasize that still the concept of our structure needs to be fully validated by experimental results. However, we also note that measurement results of many circuit-based metasurfaces containing op-amps have been so far reported in the MHz band. For instance, recently studied non-Foster-loaded metasurfaces have been experimentally validated in the MHz band\cite{nonFoster}. The maximum frequency to apply the concept of the PRC-based waveform-selective metasurface depends on what PRCs or op-amps are used, which is potentially more improved by using latest semiconductor technologies but was outside the scope of this study as an important point here was that the operating power level of waveform-selective metasurface can be improved by lowering the turn-on voltage of rectifier circuits used. Additionally, we note again that introducing lumped circuit elements (e.g., lumped capacitors) to the gap between conducting patches leads to readily lowering the operating frequency\cite{wakatsuchi2015waveformJAP}, which implies that our structure maintains the same operating frequency with a smaller periodicity. The only reason to adopt large physical dimensions in this study lies in avoiding a complexity in the structure and clarifying its underlying mechanism in a simple manner. 

As a final remark, to theoretically predict a receiver power in wireless communication systems, propagation models have been so far discussed, i.e., two-wave propagation and multipath propagation models \cite{goldsmith2005wireless}. Based on a simple two-wave propagation model with a transmitting power set to 0 dBm, for instance, an antenna located within 10 metres or so receives a signal level larger than -30 dBm. This power level is satisfied by our proposed approach. Although there still remain some issues including experimental validation, our results therefore contribute to exploiting the concept of waveform selectivity not only in fundamental electromagnetic research field but also in a more practical field of wireless communications.  

\section*{Conclusion}
We have proposed a new type of waveform-selective metasurface composed of PRCs instead of conventional schottky diodes. Our simulation results demonstrated that with proper design parameters applied to integrated resistors, PRCs had turn-on voltage that was inversely proportional to the gain of op-amps and lower than that of conventional schottky diodes. By using these PRCs, waveform-selective metasurfaces successfully reduced operating power level to approximately a thousandth, which was still changeable depending on the gain of op-amps used. As a result, waveform-selective absorption was achieved at a power level of -30 dBm, which is the same level as general communication systems, for example, Wi-Fi systems \cite{goldsmith2005wireless, IEEE80211WLAN2016}. In addition, the proposed structure exhibited a markedly wide dynamic range from -30 to 6 dBm, compared to a conventional structure that operated only around 0 dBm. Hence, our study is expected to open up the door to exploit the concept of waveform selectivity not only in fundamental electromagnetic research field but also in a more practical field of wireless communications to control different small signals at the same frequency \cite{ushikoshi2019experimental}. 

\providecommand{\noopsort}[1]{}\providecommand{\singleletter}[1]{#1}%

\section*{Acknowledgements}

This work was supported in part by Denso, SOKEN and the Support Center for Advanced Telecommunications Technology Research (SCAT).

\section*{Author contributions statement}

H.W. designed the entire project. M.T. designed specific PRCs and performed numerical simulations. All the authors considered the results and contributed to writing up the manuscript.  

\section*{Competing financial interests}
The authors declare no competing interests. 

\end{document}